\documentclass[12pt,twoside]{article}
\usepackage{fleqn,espcrc1}
\usepackage{epsfig}

\newcommand{\AmS}{{\protect\the\textfont2
  A\kern-.1667em\lower.5ex\hbox{M}\kern-.125emS}}

\title{Scalar Meson Decay Constants and the Nature of the $a_0(980)$}

\author{K. Maltman\address{Dept. Mathematics and Statistics, York Univ.,
        4700 Keele St., Toronto, ON Canada, \\ 
        and CSSM, Univ. of Adelaide, Adelaide, SA
        Australia}}%
       
\begin{document}

\maketitle

\begin{abstract}
The $a_0(980)$, $a_0(1450)$ and $K_0^*(1430)$ decay constants
are determined using a form of QCD sum rules known
to produce a very accurate determination of the $\rho$ decay
constant.  The ratio of $a_0(980)$ to $K_0^*(1430)$ decay constants
is shown to be $\sim 0.6$, ruling out both the
``loosely-bound-$K\bar{K}$-molecule'' and Gribov minion scenarios
for the $a_0(980)$.
Solutions for the isovector scalar spectral function obtained 
in the literature from sum rule analyses employing a more restrictive 
single-resonance-plus-continuum form
of the input spectral ansatz, are also investigated.  These solutions,
which suggest, in contrast to
the present results, negligible coupling of the 
$a_0(980)$ to the isovector scalar density
are shown to produce a very poor match between the OPE and
hadronic sides of the sum rules employed here, and hence to be ruled out.
\end{abstract}

\section{Introduction}

Scenarios proposed in the literature for the nature 
of the $f_0(980)$ and $a_0(980)$ 
(the loosely-bound $K\bar{K}$ molecule, crypto-exotic (four-quark), 
unitarized quark model (UQM), 
and Gribov minion pictures) differ
significantly in their spatial extent.  Processes 
previously proposed to distinguish between these different
scenarios (the $\gamma\gamma$ decay widths, and
$\phi\rightarrow\gamma a_0,f_0$) suffer
from difficult-to-quantify theoretical
uncertainties associated with the necessity of modelling
the non-trivial dynamics of the processes in question.  In this paper we
show how to determine the scalar decay constants of the scalar mesons
and use this information to make progress in distinguishing
between the different scenarios.

Because the various scenarios
differ significantly in their spatial extent, pointlike probes
such as decay constants are ideal for distinguishing amongst them.
Relations between decay constants often provide non-trivial information 
on $SU(3)_F$
classification and/or mixing. 
For example, the scenario in which
the $\pi$ and $K$, though having $m_K/m_\pi\simeq 4.5$,
are assigned to the same pseudo-Goldstone boson octet 
{\it requires} the approximate equality of the $\pi$ and $K$ decay
constants, as observed experimentally.  Similarly, 
$f_{K^*}=1.1 f_\rho\simeq f_\rho$ confirms the assignment
of the $\rho$ and $K^*$ to the same $SU(3)_F$ multiplet,
while $f^{EM}_\omega\simeq f^{EM}_\rho /3$,
(rather than $f^{EM}_\omega\simeq f^{EM}_\rho /\sqrt{3}$, as
expected for a pure octet $\omega$) confirms the 
near-ideal mixing of the vector meson sector.
In this paper, the basic idea is to determine the 
various scalar meson decay constants 
and, using the ``normal quark model state'' $K_0^*(1430)$ for reference,
investigate which (if either) of the known $a_0$
resonances might, given the values of {\it their}
decay constants, belong to the same $SU(3)_F$ multiplet.

\section{Determining The Scalar Meson Decay Constants}
We first determine the decay constant of the 
``reference'' quark model state, the $K_0^*(1430)$.
With $J_{us}=(m_s-m_u)\bar{s}u$,
we define $f_{K_0^*}$ via
$\langle 0\vert (m_s-m_u)\bar{s}u\vert K^+\rangle = f_{K_0^*}m_{K_0^*}^2$.
$f_{K_0^*}$ can be read off from the $K_0^*(1430)$ peak value
of the spectral function, $\rho_{us}$, of the corresponding correlator,
$\Pi_{us} (q^2) \equiv\ i\int\, dx e^{iq\cdot x}\langle 0\vert
T\left( J_{us}(x)J_{us}^\dagger (0)\right)\vert 0\rangle$.
Since s-wave
$K\pi$ scattering is elastic up through the
$K_0^*(1430)$\cite{LASS}, the spectral
function is saturated by $K\pi$ intermediate states out to 
$s\equiv q^2\sim 2$ GeV$^2$.
Unitarity then allows the spectral function to be
expressed in terms of the timelike scalar $K\pi$ form factor, $f_{K\pi}(s)$.
$f_{K\pi}$, in turn, satisfies an Omnes relation whose overall
normalization is set by $K_{e3}$ data, and whose phase (appearing
in the integral which defines the Omnes function) is, up to the
onset of inelasticity, simply the $I=1/2$ $K\pi$ scattering phase
shift\cite{jm,cfnp}.  At the largest $s$ for which it has been measured
($\simeq 2.9$ GeV$^2$), the $K\pi$ phase
has essentially reached its known asymptotic value, $\pi$. By
assuming (1) the absence of a possible polynomial prefactor in the
Omnes relation and (2) that the phase is $\pi$ from $2.9$ GeV$^2$
to $\infty$, one can thus construct the $K\pi$ part of $\rho_{us}$,
and hence determine $f_{K_0^*}$.  The result of this exercise is
\begin{equation}
f_{K_0^*}m^2_{K_0^*}=0.0842\pm 0.0045\ {GeV}^3\ .
\label{k01430}
\end{equation}
The input theoretical assumptions are supported by the following
observations:
(1) the determination
of $m_s$ associated with a finite energy sum rule (FESR) analysis
of $\Pi_{us}$, using $\rho_{us}$ as
generated above, is extremely stable, and produces an extremely
good match between hadronic and OPE sides\cite{kmssms}; 
(2) the resulting $m_s$ value is reproduced by a recent analysis
based on flavor breaking in hadronic $\tau$ decays (which involves
NO such additional theoretical assumptions)\cite{kmjk00}.

To determine the $a_0(980)$ and $a_0(1450)$ decay constants,
we employ a form of FESR tested in the isovector
vector channel and shown to produce 
a determination of the $\rho$ decay constant, {\it using
only OPE information, with experimental $\alpha_s$ values as the
dominant input}, accurate to within experimental errors\cite{kmfesr,kma0}.
The general FESR relation for a typical correlator $\Pi$ is
$\int_{s_{th}}^{s_0}\, ds\, w(s){\rho (s)}={\frac{-1}{2\pi i}}
\oint_{\vert s\vert =s_0}\, ds\, w(s)\Pi (s)$,
with $w(s)$ any function analytic in the region of the contour, $s_{th}$
the physical threshold, and $\rho (s)$ the corresponding spectral
function.  $s_0$ is to be chosen large enough
that the OPE can be reliably employed on the RHS.
Weight functions satisfying $w(s_0)=0$, which cut out
the region of the integral over the circle near the timelike real
axis, have been shown to produce sum rules very well satisfied, even down
to (surprisingly) low scales $s_0\sim 2$ GeV$^2$\cite{kmfesr}.

Since, in the isovector scalar channel, the $a_0(980)$ and $a_0(1450)$
are well separated, it is sufficient to take for the hadronic
spectral ansatz an incoherent sum of two Breit-Wigner resonance forms
using PDG values for the masses and widths.
The decay constants are to be fit
in the FESR analysis, which works by using analyticity,
together with qualitative non-perturbative
input (the known resonance positions and widths),
to essentially ``measure'' 
the decay constants in terms of $\alpha_s$ (the $D=0$ OPE terms 
dominate at scales $s>2$ GeV$^2$).
We work with the correlator of the
scalar density $J_{ud}\equiv (m_s-m_u)\bar{d}u$, the mass factor being chosen
so as to cancel in the ratio of
$a_0$ to $K_0^*(1430)$ decay constants.
The ratio then reduces to that of the matrix elements of
the $\bar{d}u$ and $\bar{s}u$ densities which, since these densities
are members of the same $SU(3)_F$ octet, must reduce to
$1$ in the $SU(3)_F$ limit for an $a_0$ lying
in the same multiplet as the $K_0^*(1430)$.  

On the OPE side of our FESR's, the dominant
$D=0$ part of the OPE is known to four-loop order\cite{jm,chetyrkin}, and
the small higher $D$ terms are also known out to $D=6$\cite{jm}.
Instanton contributions are determined using the instanton
liquid model\cite{ilm}.  A more detailed description of the OPE
input and the method
of calculation can be found in Ref.~\cite{kma0}.

Fitting the $a_0$ decay constants using the OPE as described above,
one finds 
\begin{eqnarray}
f_{a_0}m^2_{a_0}&=&0.0447\pm 0.0085\ {GeV}^3\ ,  \nonumber\\
f_{a_0^\prime}m^2_{a_0^\prime}&=&0.0647\pm 0.0123\ {GeV}^3\ .
\label{a0results}
\end{eqnarray}
The errors are dominated by the estimate of the
uncertainty associated with truncating the dominant $D=0$
part of the OPE at 4-loop order.
The quality of agreement between the OPE and hadronic side
which results is shown in Figure 1 for the weight choice
$w(s)=\left( 1-s/s_0\right)\left( 2-s/s_0\right)$ (chosen to
reduce the sensitivity to the less-well-known instanton
contributions).  The dotted line is the OPE
side of the sum rule and the dashed-dotted line the hadronic
side obtained using the results of Eqs. (\ref{a0results}).
If the $a_0(980)$ is very diffuse compared to $K_0^*(1430)$
(the loosely-bound $K\bar{K}$ molecule scenario), 
one should find a much smaller decay constant;
if very compact (the minion scenario), 
a much larger decay constant.
The results of Eqs. (\ref{a0results}) rule out both of these
scenarios.  Two additional possibilities need to be considered in 
more detail to make this conclusion definitive
in the molecule case.  The $a_0(980)$ spectral strength
is proportional to the square of the decay constant.
To see if this small-decay-constant scenario
is plausible, we set 
the coefficient of the $a_0(980)$ Breit-Wigner to zero
by hand and re-optimize the $a_0(1450)$ decay constant.
The best fit obtained from this exercise is shown by the solid
line in Figure 1; the match to the OPE side is clearly terrible.
This failure cannot be cured
by using a broad background, rather than narrow
resonance contribution, in the
region below the $a_0(1450)$.  Indeed,
the $\pi\eta$ matrix element
of the scalar density can be computed unambiguously
to leading order in
the chiral expansion using Chiral Perturbation
Theory (ChPT).  Given this matrix element,
the corresponding background contribution to the spectral function
can be obtained using unitarity.
Setting the $a_0(980)$ resonance contribution to zero, and
optimizing the $a_0(1450)$ decay constant in the
presence of the resulting background, one obtains a ``best''
fit almost
identical to that given by the solid line.  Multiplying the ChPT-generated
background contribution by a factor of $5$ to allow (generously)
for higher (chiral) order contributions, one obtains the ``best'' fit
shown by the dashed line.  Clearly no
version of the loosely-bound molecule scenario corresponds
to a good match to the OPE side, thus ruling
out this scenario.  The relation between the $a_0(980)$ and
$K_0^*(1430)$ decay constants given by the results above is,
in contrast, exactly what one would expect in the UQM scenario
if the $a_0(980)$ were a roughly equal admixture of a normal
quark model meson core and a loosely bound two-meson component.
Because of the additional hidden strange pair present in the $a_0(980)$ in
the cryptoexotic scenario, a natural expectation would be to find
an $a_0(980)$ 
decay constant significantly smaller than that of the $K_0^*(1430)$.
A calculation of the decay constant
in this scenario would, however, be welcome and,
since
the results above represent an (albeit indirect)
``measurement'' of the scalar meson decay constants 
(basically in terms of $\alpha_s$,
which is very well known at the scales in question),
would serve to
provide a further, highly non-trivial test of the 
cryptoexotic scenario. 

\begin{figure}[htb]
\begin{minipage}[t]{75mm}
\caption{OPE and hadronic sides of the isovector scalar sum rule
for the various spectral ansatze discussed in the text.}
\label{fig:1}
\psfig{figure=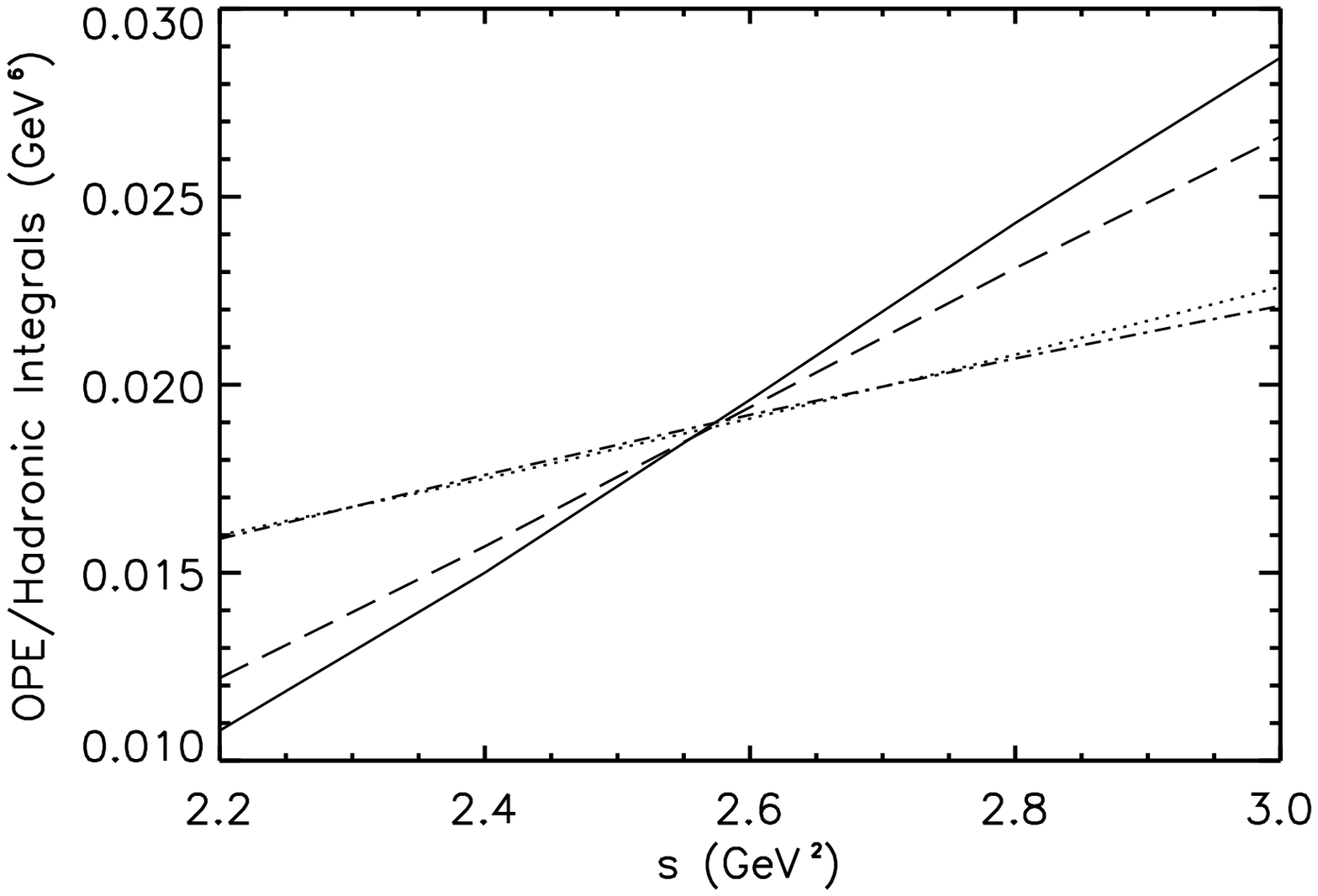,height=2.4in,width=2.8in}

\end{minipage}
\hspace{\fill}
\begin{minipage}[t]{75mm}
\caption{Testing the ``best fit'' spectral solution of Ref.\cite{elias}}

\label{fig:2}

\psfig{figure=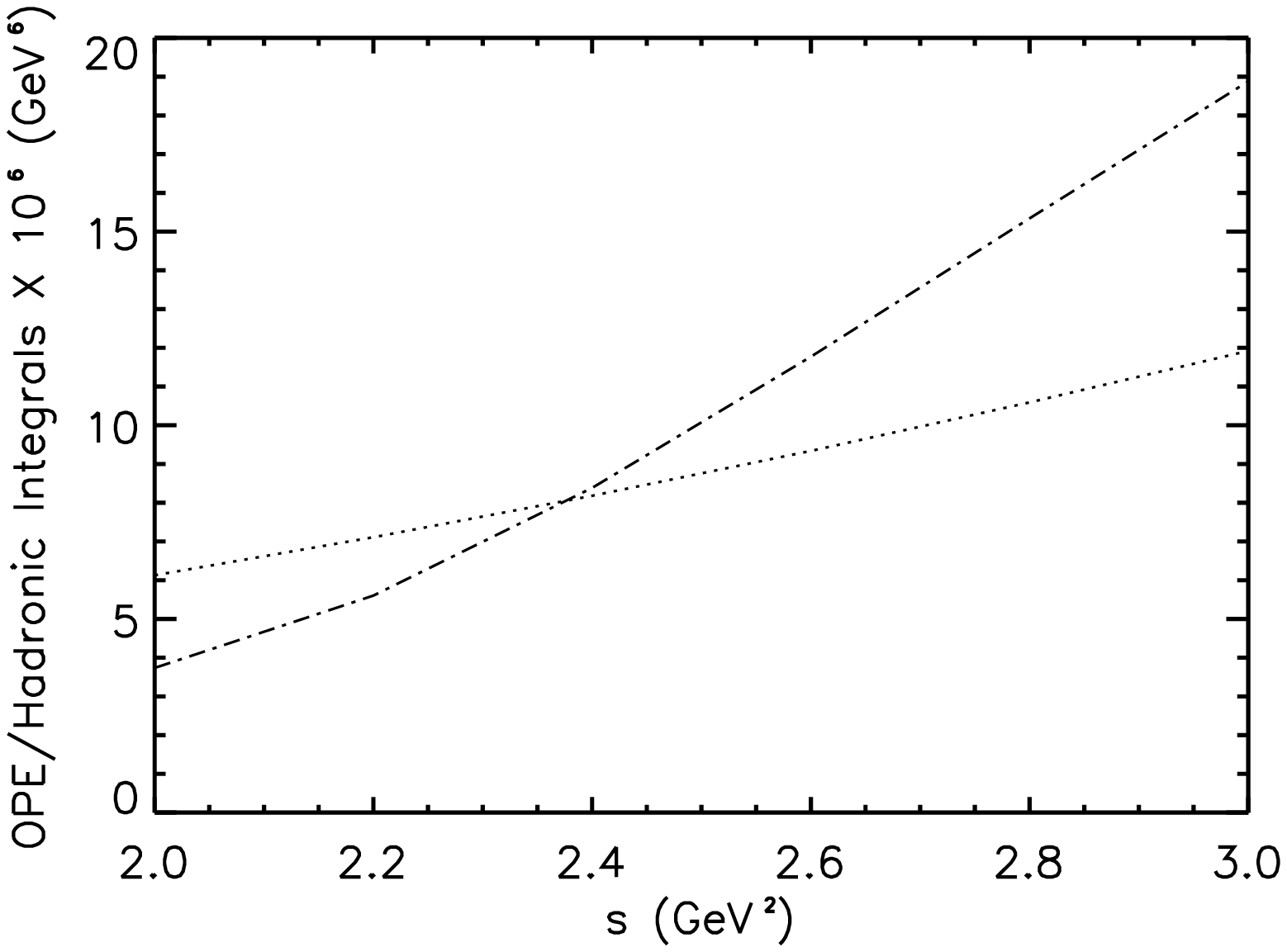,height=2.4in,width=2.8in}
\end{minipage}
\end{figure}

We conclude by discussing a recent claim that the 
$a_0(980)$ coupling to
$J^\prime_{ud}=(m_u+m_d)\bar{d}u$
is small, and that the spectral distribution is dominated by
a contribution with $m\simeq 1.5$ GeV\cite{elias}.  This claim
(which clearly conflicts with the results above) is based on
a Laplace sum rule analysis of the $J^\prime_{ud}$
correlator (which differs from the correlator considered
above by the overall multiplicative factor $[(m_u+m_d)/(m_s-m_u)]^2$)
{\it assuming a form for the spectral function
consisting of a single resonance plus an OPE-generated ``continuum'' 
beyond some ``continuum threshold'', $s_0$}.
The resonance mass and $s_0$ are fit in 
the sum rule analysis, whose validity,
apart from the question of the suitability of the form of the
spectral ansatz, relies only on analyticity and
the applicability of the OPE, as in the sum rules above.  {\it IF}
these assumptions are valid, and {\it IF} the spectral function
resulting from the fitting procedure is physical, then
FESR's analogous to those above must also be valid.  Testing the
spectral solution of Ref.~\cite{elias} 
by means of the resulting FESR, 
one finds the results shown in Figure 2.
The dotted line again represents the OPE side,
and the dashed-dotted line the hadronic side, of the FESR.
One immediately sees that,
although the solution of Ref.~\cite{elias}
may represent the ``best'' fit within the restricted form of the
spectral ansatz employed, the quality of the OPE/hadronic match
is, in fact, very poor, and, moreover,
far inferior to that of the 
two-resonance form discussed above.  The results of Ref.~\cite{elias},
and the apparent contradiction with the results obtained here
would, therefore, appear to be
the artifact of the use of an overly-restrictive form for the
the spectral ansatz.

\end{document}